\documentstyle[12pt]{article}
\begin{document}
\title{Magnetic flux locking in two weakly coupled superconducting rings}
\author{R.de Bruyn \ Ouboter~${}^{a}$  , A.N.\ Omelyanchouk~${}^{b}$ , E.D.\ Vol~${}^{b}$ 
\\ 
{\small {\em ${}^{a}$ Kamerlingh Onnes Laboratory,}}
\\
{\small {\em Leiden Institute of Physics,}}
\\
{\small {\em Leiden University,}}
\\
{\small {\em P.O. 9506, 2300 RA Leiden, The Netherlands}} 
\\
{\small {\em ${}^{b}$ B.Verkin Institute for Low Temperature Physics and
Engineering,}}
 \\
{\small {\em National Academy of Sciences of Ukraine,}}
\\
{\small {\em 47 Lenin Ave., 310164 Kharkov, Ukraine}}}
\date{}
\maketitle
\begin{abstract}
We have analyzed the quantum interference effects in the macroscopic
''superconducting molecule''. The composite system consists of two massive
superconducting rings, each interrupted by a Josephson junction, which are
at the same time weakly coupled with one another. The special case of
coupling via the Josephson four-terminal junction is considered. The
structure of the macroscopic quantum states in an applied magnetic field is
calculated. It is shown, that depending on the values of the magnetic fluxes
through each ring, the system displays two groups of states, the
''orthostates'' with both induced currents going in the same direction, and
the ''parastates'' with the opposite currents and with the total induced
flux locked to zero value. The transition to the flux locked state with
changing of the total applied flux is sudden and is preserved in a certain
interval which is determined by the difference of the fluxes applied through
each ring. It makes the system sensitive to small gradients of the external
magnetic field.

\end{abstract}
\newpage
The system which we studied, is shown in Fig.1 and consists of two bulk
superconducting rings, coupled via the 4-terminal Josephson junction
\cite{vo1,oov97}. The 4-terminal Josephson junction is a system of two
microbridges, $1-2$ and $3-4$, having the common centre $"o"$. The
interference in the cross section $"o"$ of macroscopic wave functions $\Psi
_j$ of the $j$th terminal ($j=1,..4$) leads to nonlinear coupling and
consequently to interference between the current states in each ring. The
resulting current state of the whole system can be regulated by the
difference of the magnetic fluxes applied through the rings, in analogy with
the phase difference between two weakly coupled bulk superconductors. The
studying of the macroscopic quantum states of such ''superconducting
molecule'' is the subject of the present paper.

The free energy $U$ of our system in an applied magnetic field contains the
magnetic energy $U_m$ and the Josephson coupling energy $U_J$. The energy $%
U_m$ has the form \cite{lali}:

\begin{equation}
U_m=\frac{(\Phi _1^e-\Phi _1)^2L_2}{2(L_1L_2-L_{12}^2)}+\frac{(\Phi
_2^e-\Phi _2)^2L_1}{2(L_1L_2-L_{12}^2)}-\frac{L_{12}}{(L_1L_2-L_{12}^2)}%
(\Phi _1^e-\Phi _1)(\Phi _2^e-\Phi _2), 
\end{equation}
where $\Phi _{1,2}^e$ are the external magnetic fluxes applied to the rings $%
1,2$ and $\Phi _{1,2}$ are the resulting fluxes embraced by the rings; $%
L_{1,2}$ and $L_{12}$ are the ring self-inductances and the mutual
inductance ($L_{12}^2<L_1L_2$) . The coupling energy $U_J$ (in dimensionless
units) is expressed in terms of phases $\varphi _j$ ($j=1,..4)$ of the
superconducting order parameter in the $j$th terminal \cite{oov97}:

\begin{equation}
U_J=-\kappa ^2\cos {}^2\frac{\phi _1}2-\cos {}^2\frac{\phi _2}2-2\kappa \cos
{}\frac{\phi _1}2\cos {}\frac{\phi _2}2\cos \chi , 
\end{equation}
if we introduce the phase differences across the weak links in the rings

$$
\phi _1=\varphi _1-\varphi _2,\phi _2=\varphi _3-\varphi _4 
$$
and the ''total'' phase difference between the rings

$$
\chi =\frac{\varphi _1+\varphi _2}2-\frac{\varphi _3+\varphi _4}2. 
$$
The coupling constant $\kappa $ in (2) is the ratio of critical currents of
the weak links $1-2$ and $3-4$. In the following for simplicity we will
consider the case of identical rings with $L_1=L_2=L$ and the symmetrical
coupling\ $\kappa =1$ ($I_{c,12}=I_{c,34}=I_c$).

The phase differences $\phi _{1,2}$ are related to the magnetic fluxes $\Phi
_{1,2}$ by : $\varphi _{1,2}=-2e\Phi _{1,2}/\hbar $. Thus, the total energy
in reduced units of the two coupled rings as function of the embraced
magnetic fluxes at given values of the applied fluxes is defined as

\begin{equation}
\begin{array}{r}
U(\Phi _1,\Phi _2,\chi \mid \Phi _1^e,\Phi _2^e)= 
\frac{(\Phi _1^e-\Phi _1)^2}{2{\cal L}}+\frac{(\Phi _2^e-\Phi _2)^2}{2{\cal L%
}}-\frac \ell {{\cal L}}(\Phi _1^e-\Phi _1)(\Phi _2^e-\Phi _2)- \\  \\ 
-\cos {}^2\frac{\Phi _1}2-\cos {}^2\frac{\Phi _2}2-2\cos {}\frac{\Phi _1}%
2\cos {}\frac{\Phi _2}2\cos \chi , 
\end{array}
\end{equation}
where $\ell =L_{12}/L$ the normalized mutual inductance ($\ell <1),{\cal L}~
=(2eI_c/\hbar )L(1-\ell ^2)$ the dimensionless effective self-inductance;
the magnetic fluxes are measured in units $\hbar /2e$. Note the dependence
of the potential $U$ on the phase $\chi $. As we will see, in the stable
steady state the phase $\chi $ can take only the value $0$ or $\pi $, which
corresponds to existence of two groups of states with different symmetry.

The minima of the potential $U$ (3) with respect to variables $\Phi _1$, $%
\Phi _2$, $\chi $ at given external fluxes $\Phi _1^e$ and $\Phi _2^e$
determine the stable steady states of our system. The minimization of $U$
with respect to $\chi $ gives that the phase $\chi $ takes the value $0$ or $%
\pi $, depending on the equilibrium values of $\Phi _1$and $\Phi _2$

\begin{equation}
\cos \chi =sign(\cos \frac{\Phi _1}2\cos \frac{\Phi _2}2). 
\end{equation}

In the steady state $\frac{\partial U}{\partial \Phi _1}=\frac{\partial U}{%
\partial \Phi _2}=0$, or : 
\begin{equation}
\Phi _1^e-\ell \Phi _2^e=\Phi _1-\ell \Phi _2+{\cal L\sin }\frac{\Phi _1}%
2[\cos \frac{\Phi _1}2+\cos \chi \cos \frac{\Phi _2}2], 
\end{equation}

\begin{equation}
\Phi _2^e-\ell \Phi _1^e=\Phi _2-\ell \Phi _1+{\cal L}\sin \frac{\Phi _2}%
2[\cos \frac{\Phi _2}2+\cos \chi \cos \frac{\Phi _1}2)], 
\end{equation}
with $\cos \chi $ defined by the condition (4).

The solutions of eqs.(5) and (6) $\{\Phi _1,\Phi _2\}$ which correspond to
the minima of the potential $U$ must satisfy the requirements:

\begin{equation}
\frac{\partial ^2U}{\partial \Phi _1^2}>0,\frac{\partial ^2U}{\partial \Phi
_2^2}>0,\frac{\partial ^2U}{\partial \Phi _1^2}\frac{\partial ^2U}{\partial
\Phi _2^2}-(\frac{\partial ^2U}{\partial \Phi _1\partial \Phi _2})^2>0.
\end{equation}
It can be shown that the conditions (7) are fulfilled for all values of $%
\Phi _1$and $\Phi _2$ if ${\cal L}+\ell <1$. In the following we will
consider the case when the inductances ${\cal L}$ and $\ell $ are small
enough to satisfy this inequality. Thus, all solutions $\{\Phi _1,\Phi _2\}$
of the equations (4), (5) and (6) determine the possible stable or
metastable states of the system. The circulating ringcurrents $I_{1,2}$ in
state $\{\Phi _1,\Phi _2\}$ are:

\begin{equation}
I_1=-\frac 12\sin \Phi _1-\sin \frac{\Phi _1}2sign(\cos \frac{\Phi _1}2)\mid
\cos \frac{\Phi _2}2\mid , 
\end{equation}

\begin{equation}
I_2=-\frac 12\sin \Phi _2-\sin \frac{\Phi _2}2sign(\cos \frac{\Phi _2}2)\mid
\cos \frac{\Phi _1}2\mid 
\end{equation}
in units of $I_c$.

The value of $\cos \chi $ in eqs.(5), (6), which equals $\pm 1$, determines
two possible ''binding'' of the current states in individual rings. The
group of states $\{\Phi _1,\Phi _2,\chi =0\}$ we will call symmetric, or
''ortho'', states and the group of states $\{\Phi _1,\Phi _2,\chi =\pi \}$ -
antisymmetric, or ''para'', states. As we will see, the first one
corresponds to the induced ringcurrents going in the same direction, and the
second one - to the currents going opposite.

We will study the behaviour of our system in an applied magnetic field as
the response on the total applied magnetic flux $\Phi ^e=\Phi _1^e+\Phi _2^e$
at given difference $\delta ^e=\Phi _1^e-\Phi _2^e$ of the fluxes through
the each ring. The state of the system as whole is determined by the total
embraced magnetic flux $\Phi =\Phi _1+\Phi _2$ , or by the total orbital
magnetic moment $M$, which is proportional to the sum of the induced
ringcurrents , $I=I_1+I_2$. Note, that the positive (negative) sign of $I$
corresponds to the parallel (antiparallel) direction of $M$ with respect to
the external magnetic field $H$. From the (4-6) we obtain:

\begin{equation}
\Phi _e=\Phi +\frac{{\cal L}}{1-\ell }\sin \frac \Phi 2[\cos \frac \delta
2+\cos \chi ], 
\end{equation}

\begin{equation}
\delta ^e=\delta +\frac{{\cal L}}{1+\ell }\sin \frac \delta 2[\cos \frac
\Phi 2+\cos \chi ], 
\end{equation}
\begin{equation}
\cos \chi =sign(\cos \frac \Phi 2+\cos \frac \delta 2), 
\end{equation}
where $\delta $$=\Phi _1-\Phi _2$.

Let us start from the case of small inductances $\ell ,{\cal L}\ll 1.$ If $%
\delta ^e=0$, from the eqs.(11), (12) follows that $\delta =0$ and $\chi =0$%
. For $\Phi (\Phi ^e)$ we have the usual equation $\Phi _e=\Phi +2{\cal L}%
\sin \frac \Phi 2$ for the case of decoupled rings \cite{bar}, each interrupted
by a Josephson junction. At $\delta ^e\neq 0$ and consequently $\delta \neq 0
$, the solutions with $\chi =\pi $ are possible in the vicinity of $\Phi
\approx 2\pi $. In the limit ${\cal L}\rightarrow 0$ for the total induced
magnetic flux $\Phi (\Phi ^e,\delta ^e)$ we have the expression

\begin{equation}
\Phi =\Phi ^e-{\cal L}\sin \frac{\Phi ^e}2[\cos \frac{\delta ^e}2+sign(\cos 
\frac{\Phi ^e}2+\cos \frac{\delta ^e}2)]. 
\end{equation}
In the case of small $\delta ^e\ll 1$ it becomes :
\begin{equation}
\Phi = \left\{ \begin{array}{ll}
\Phi ^e-2{\cal L}\sin \frac{\Phi ^e}2 & \mbox{if $ \mid \Phi ^e - 2\pi \mid > \mid \delta ^e \mid $ }\\
\Phi ^e & \mbox{if $ \mid \Phi ^e - 2\pi \mid < \mid \delta ^e\mid $. }
\end{array}
\right.
\end{equation}
Thus, for given value of $\delta ^e$ with changing of the total applied flux 
$\Phi ^e$ the system switches from the state with $\chi =0$ to the state
with $\chi =\pi $. In interval $2\pi -\delta ^e<\Phi ^e<2\pi +\delta ^e$ the
total induced flux $\Phi -\Phi ^e$ equals to zero for $\delta ^e\ll 1$. We
call such behaviour magnetic flux locking. It is emphasized that the
transition to the flux locked state is sudden and is preserved in a certain
interval of the applied magnetic flux. For the sum of the induced
ringcurrents $I=I_1+I_2$, in the limit ${\cal L}\rightarrow 0$ we have

\begin{equation}
I(\Phi ^e,\delta ^e)=-\sin \frac{\Phi ^e}2[\cos \frac{\delta ^e}2+sign(\cos 
\frac{\Phi ^e}2+\cos \frac{\delta ^e}2)].
\end{equation}
In the flux locked state the total current $I$ equals to zero in
correspondence with (14). Thus the ringcurrents $I_{1,2}$ are going in
opposite directions and compensate each other, or the system is in the
''para'' state . The complete compensation takes place for $\delta ^e\ll 1$,
with the corrections to zero value being of the order of $(\delta ^e)^2$. In
Fig.2 we plot the dependence of $I(\Phi ^e)$ (15) for the flux difference $%
\delta ^e=2{\pi}/10$, or in dimension units $1/10$ of a flux quantum $h/2e$.
The dashed line is the sum of the currents in the same, but
decoupled, rings with the same applied fluxes $\Phi _1^e=1/2(\Phi ^e+\delta
^e)$ and $\Phi _2^e=1/2(\Phi ^e-\delta ^e)$ . The magnetic susceptibility of
the system as whole is proportional to $-\frac{\partial I}{\partial \Phi ^e}$
and will reflect the behaviour of the induced currents.

For finite, but small, values of the inductances , the behaviour described
above will be qualitatively the same. Only instead of the sharp switches
hysteretic regions appear, of which the width  is proportional to ${\cal L\ }
$. In Fig.3 the dependence $\Phi (\Phi ^e)$ for ${\cal L}$ $=0.25,\ell =0$
and $\delta ^e=1$ is shown, as follows from the numerical solution of the
eqs.(10-12). Naturally, these hysteretic regions will be smeared by thermal
fluctuations (see the analysis of the influence of noise on the similar
system, the so called 4-terminal SQUID, in ref.\cite{oo98}).

In conclusion, we have studied the macroscopic quantum states in the system
of two weakly coupled superconducting rings.The nonlinear coupling leads to
interference between the current states in each ring. It is manifested as
the cooperative behaviour of the rings in some region of the applied magnetic
fluxes, which we call magnetic flux locking. We would like to remark that
our macroscopic approach is not restricted by the special kind of the
coupling through the crossed superconducting bridges. In fact, any
mesoscopic 4-terminal weak link will produce a coupling similar to the $U_J$
(2). For example, it can be the experimental setup described in ref.\cite{heid},
namely the two-dimensional normal layer which is connected with four
terminals instead of the two ones as studied in ref.\cite{heid}. 

One of the authors, A.N.O., would like to acknowledge the support for this
research from the Kamerlingh Onnes Laboratory, Leiden University.

\newpage

FIGURE CAPTIONS \ \ \ \

Figure 1. \ \ The two bulk superconducting rings, coupled via the 4-terminal
Josephson junction ( the region closed by the dashed lines, of which the area is  
of the order of the coherence length squared). \ \

Figure 2.  \ \ The total induced current as a function of the total applied flux
at given difference of applied fluxes through each ring ${\delta}^e=2{\pi}/10$ (or $1/10$ of a flux
quantum $h/2e$). ${\cal L}=0$. The dashed line is the corresponding dependence 
in the case of decoupled rings.\ \

Figure 3. \ \ The dependence of the total magnetic flux ${\Phi}$ on the total external
flux ${\Phi}^e$ for ${\delta}^e=1, {\cal L}=0.25, {\ell}=0$. The arrows indicate the jumps of
the flux from metastable to stable states. The dashed line is ${\Phi}={\Phi}^e$. \ \

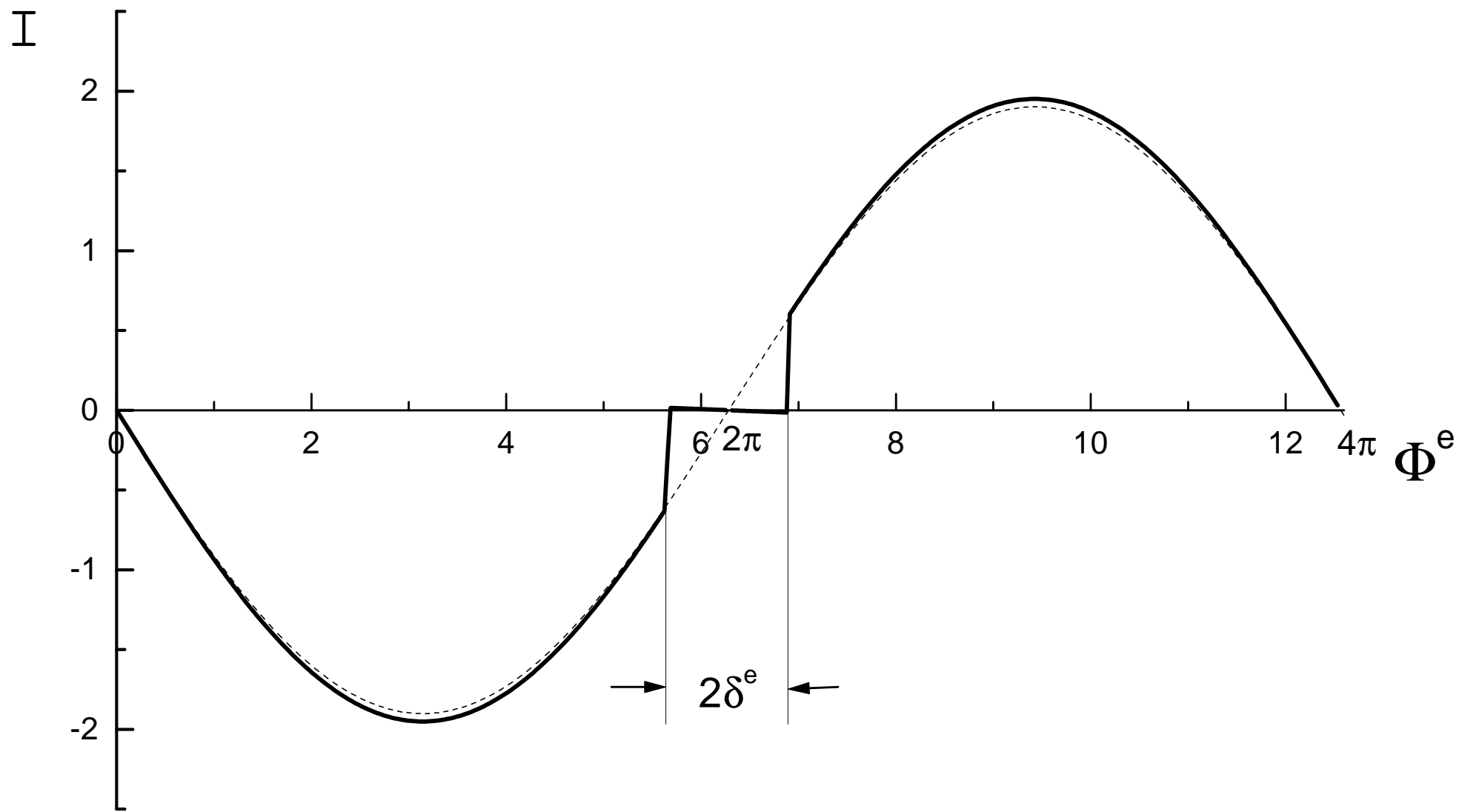
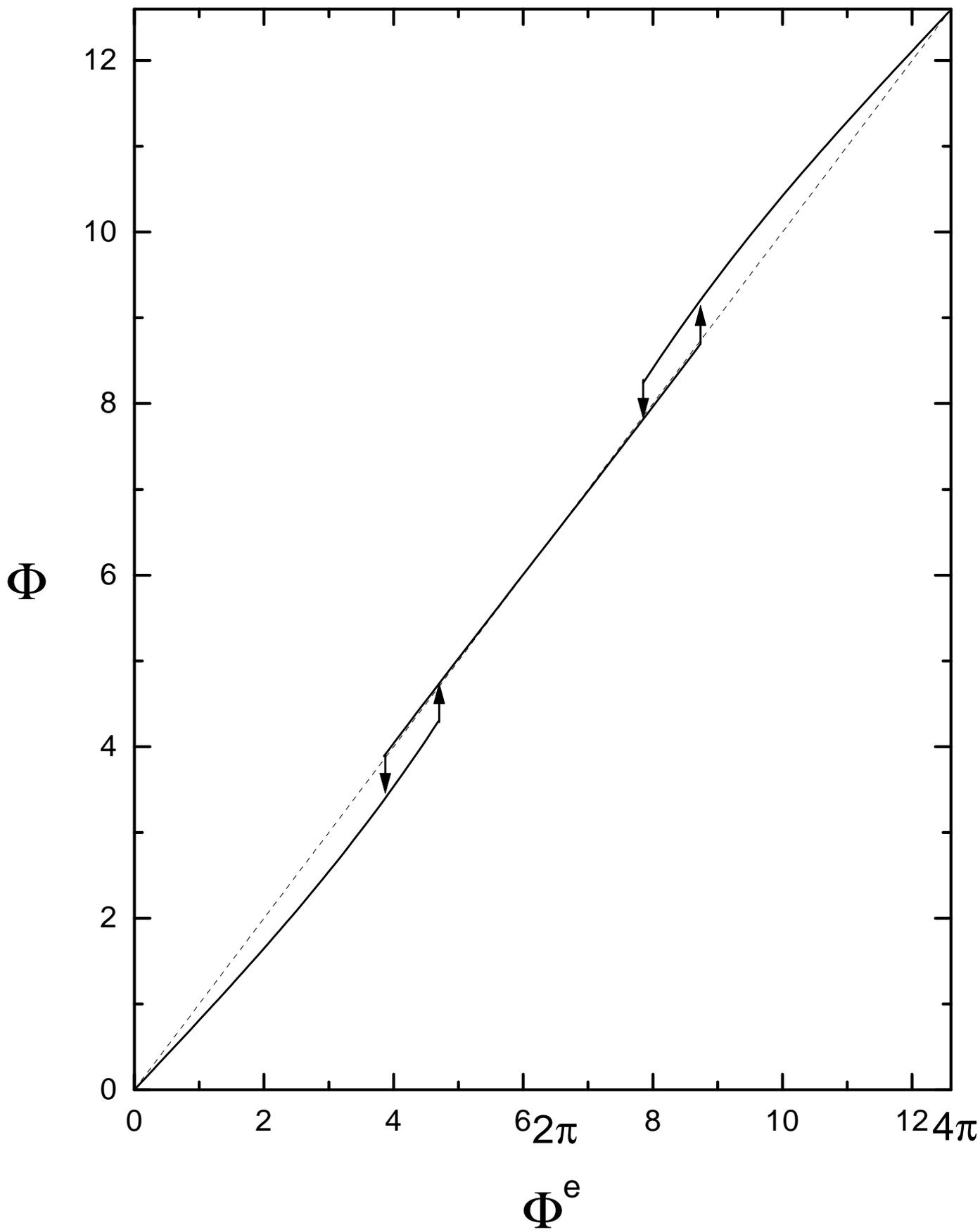
\end{document}